# Good Practice Guide on the electrical characterisation of graphene using contact methods

**Edited by:**
Alexandra Fabricius (VDE)
Alessandro Catanzaro (NPL)
Alessandro Cultrera (INRIM)

**Contributions from:**
Alessandro Cultrera (INRIM)
Félix Isidro Raso Alonso (CEM)
Laura Matias Hernandez (CEM)
Alessandro Catanzaro (NPL)
Vittorio Camarchia (POLITO)
Luca Callegaro (INRIM)

**Affiliations:**
VDE, Association for Electrical, Electronic & Information Technologies, Frankfurt am Main, Germany.
NPL, National Physical Laboratory, Teddington, United Kingdom.
INRIM, Istituto Nazionale di Ricerca Metrologica, Turin, Italy.
CEM, Centro Español de Metrología, Tres Cantos, Madrid, Spain.
POLITO, Politecnico di Torino, Dipartimento di Elettronica e Telecomunicazioni, Torino, Italy.

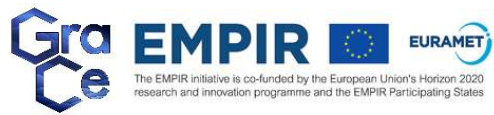

# Index

# Guide information

**What is it about?**

This guide provides protocols for determining the electrical properties of graphene sheets, on insulating substrates, using contact methods: that is by physical contacting the graphene surface with metallic electrodes. Depending on the methodology the properties that can be measured are the electrical sheet conductivity, the concentration and mobility of electrical charge carriers. Each protocol gives advice to understand the measurement principle, how to implement it to perform reliable measurements traceable to the International System of units, and hints to express the corresponding measurement uncertainty.

**Who is it for?**

The guide is for producers and users of graphene who need to understand how to measure the electrical properties of graphene. Such information is essential in a host of technology application areas where graphene may be used.

**What is its purpose?**

This guide provides a detailed description of how to determine the key electrical properties of graphene, so that the graphene community can adopt a common, metrological approach that allows the comparison of commercially available graphene materials. The guide is intended to form a bedrock for future interlaboratory comparisons and support the direct collaboration with international standardisation organisations.

**What is the pre-required knowledge?**

The guide is for users in research and industry who have experience with and access to the advanced techniques described herein. It is targeted at analytical scientists and professionals who have a Bachelor's degree in science.

# Overview

## Introduction

This guide is a deliverable of the Joint Research Project 16NRM01 GRACE --- *Developing electrical characterisation methods for future graphene electronics*. The project belongs to the European Metrology Programme for Innovation and Research (EMPIR). GRACE is framed within the Normative targeted program, and its overall goal is (1) the development of validated protocols for the measurement of the electrical properties of graphene, and their implementation in order to achieve accurate and fast-throughput measurement of graphene; and (2) the collaboration with international standardisation committees in order to initiate and develop dedicated documentary standards for the electrical characterisation of graphene. The adoption of this GPGs and, when published, the corresponding standards, will allow industry to perform accurate measurements of the electrical properties of graphene and thereby provide customers with reliable and comparable specifications of graphene as an industrial product.

## Scope

The electrical characterisation of graphene, either in plane sheets or in properly geometrised form can be approached using methods already employed for thin film materials. The extraordinary thinness (and, correspondingly, the volume) of graphene, however, makes the proper application of these methods difficult. The electrical properties of interest (sheet electrical resistivity/conductivity, concentration and mobility of charge carriers) must be indirectly derived from the measurement outcome by geometrical and electrical modelling; the assumptions behind such models (e.g., uniformity and isotropy, effective value of the applied fields, etc.) require careful consideration. The traceability of the measurement to the International System of units and a proper expression of measurement uncertainty is an issue.

This guide focuses on contact methods, that is method where the graphene sample surface is physically contacted with metallic electrodes. A companion guide about non-contact and high-throughput methods is also available.

The methods discussed are:

- the in-line four-point probe (4PP);
- the van der Pauw method (vdP) for sheet resistance measurement;
- the van der Pauw method for charge carrier mobility measurement;
- the electrical resistance tomography (ERT);
- the coplanar waveguide method (CPW).

For each method, a corresponding measurement protocol is discussed, which describes:

- the measurement principle;
- sample requirements and preparation;
- a description of the measurement equipment / apparatus;

- calibration standards and ways to achieve a traceable measurement;
- environmental conditions to be considered;
- a detailed measurement procedure, with specific hints to achieve a reliable measurement;
- modeling and data analysis to determine the electrical property of interest;
- considerations about the expression of measurement uncertainty.

## Terms and Definitions

The international standardisation organisations ISO (International Organization for Standardisation) and IEC (International Electrotechnical Commission) have published many standards in the area of measurement and characterisation of nanomaterials, particularly referring to some of the techniques detailed here. For further information the IEC/TC 113 'Nanotechnology in electrotechnical products and systems', ISO/TC 229 'Nanotechnologies', ISO/TC 24/SC 4 'Particle characterization' and ISO/TC 201 'Surface Chemical Analysis' websites should be examined, which list both the published and under-development standards in this area.

Furthermore, ISO and IEC maintain terminological databases for use in standardization:

- IEC Electropedia: available at http://www.electropedia.org/
- ISO Online browsing platform: available at http://www.iso.org/obp

In particular, the terms and definitions from "ISO/TS 80004-13: Nanotechnologies -- Vocabulary -- Part 13: Graphene and related two-dimensional (2D) materials" apply here and should be referred to.

For the purposes of this document, the following terms and definitions apply:

**Graphene; graphene layer; single layer graphene; monolayer graphene**

Single layer of carbon atoms with each atom bound to three neighbours in a honeycomb structure

*Note 1 to entry: It is an important building block of many carbon nano-objects.*

*Note 2 to entry: As graphene is a single layer, it is also sometimes called monolayer graphene or single layer graphene and abbreviated as 1LG to distinguish it from bilayer graphene (2LG) and few-layered graphene (FLG).*

*Note 3 to entry: Graphene has edges and can have defects and grain boundaries where the bonding is disrupted.*

[Source: ISO/TS 80004-13]

**Bilayer graphene (2LG)**

Two-dimensional material consisting of two well-defined stacked graphene layers

[Source: ISO/TS 80004-13]

**Few-layer graphene (FLG)**

Two-dimensional material consisting of three to ten well-defined stacked graphene layers

[Source: ISO/TS 80004-13]

**Electrical conductivity, $\sigma$**

The SI unit of measure of σ in graphene and 2D materials is the siemens, symbol S.

**Sheet resistance, $R_S$**

electrical resistance of a conductor with a square shape (width equal to length) and thickness significantly smaller than the lateral dimensions (thickness << width, length)

*Note 1 to entry: The SI unit of sheet resistance is ohms (Ω). However, for the purpose of this procedure, expressed as ohm per square (Ω/sq).*

*Note 2 to entry: We define here a "sheet resistivity" as the intrinsic, local property of two-dimensional conductor, a two-dimensional equivalent to three-dimensional resistivity. For an electrically uniform thin film material, the sheet resistance and sheet resistivity are identical.*

[Source: adapted from IEC/TS 61836:2007 Ed. 2.0]

**Drift mobility of a charge carrier, $\mu$**

the quotient of the modulus of the mean velocity of the charge carriers in the direction of an electric field by the modulus of the field strength. The SI unit of mobility is cm$^2$ / V s.

*Note 1 to entry: The (drift) mobility is here considered to be the fundamental, intrinsic (local) property. The Hall and field effect mobility are then the extrinsic (sample) electrical measurements, carried out to determine the intrinsic mobility.*

*Note 2 to entry: The (drift) mobility for electrons and holes can be very different, depending on the residual doping and scattering mechanisms for the given sample.*

[Source: adapted from IEV 521-02-58]

**Four-point probe method**

method to measure electrical sheet resistance of thin films that uses separate pairs of current-carrying and voltage-sensing electrodes

*Note 1 to entry: The method is local with a characteristic length scale defined by the probe distance, and generally requires the resistivity variations to be on a much larger scale than the probe spacing. Depending on the positions of the sample-probe contact of the four probe contacts with the surface, different geometrical factors must be used to extract the sheet resistance.*

[Source: adapted from ISO 80004-13, 5.3.1]

**Inline four-point probe method**

type of four-point probe measurement where four-point electrodes are aligned in a row

*Note 1 to entry: In this method, four probes contact the test sample in a linear arrangement. A voltage drop is measured between the two inner probes while a current source supplies current through the outer probes.*

*Note 2 to entry: The thickness of the layer needs to be small compared to its lateral dimensions so that the sample is approximately two-dimensional.*

*Note 3 to entry: The distance between the probes shall be small compared to the lateral dimensions of the sample so that edge effects on the electric field in the sample can be neglected.*

*Note 4 to entry: The resistance of the sample can be calculated by Ohm's law. Geometrical factors can be used for corrections if the sample is too small or if the measurement is performed close to the edges of the sample.*

[Source: IEC/TS 62607-6-9]

**van der Pauw method**

type of four probe measurement for samples of arbitrary shape

*Note 1 to entry: The thickness of the layer needs to be small compared to its lateral dimensions so that the sample is approximately two-dimensional*

*Note 2 to entry: The van der Pauw method requires four probes placed arbitrarily around the perimeter of the sample, in contrast to the linear four-point probe which is placed on the top of the sample.*

*Note 3 to entry: The van der Pauw method provides an average sheet resistivity of the sample.*

[Source: IEC/TS 62607-6-9]

**Electrical Resistance Tomography**

method to map the electrical conductivity of thin films that uses several contacts arranged in multiple pairs of current-carrying and voltage-sensing electrodes placed on the sample's boundary

*Note 1 to entry: electrical resistance tomography requires numerical methods in order to recover the sheet resistance of the sample from the series of boundary electrical measurements.*

**Coplanar Waveguide measurement at GHz frequency**

method to measure the propagation constant of a linear structure (the CPW) consisting in metallic ground-planes with graphene as the dielectric.

*Note 1 to entry: From the propagation constant the conductivity (and with geometrical considerations also the sheet resistance) of the sample material used placed within the CPW can be retrieved.*

# 1 Measurement of sheet resistance of graphene films with in-line 4-point probe

## 1.1 General

This part of the Good Practice Guide describes a method to determine the sheet resistance $R_S$ by the in-line four-point probe method (4PP).

The sheet resistance $R_S$ is derived by measurements of four-terminal electrical resistance performed on four electrodes placed on the surface of the planar sample.

The measurement range for the sheet resistance of the graphene layer shall be in the nominal range of 1 - $10^6$ Ω/sq.

The method is applicable for CVD graphene provided it is transferred to quartz substrates or other insulating materials as well as graphene on silicon carbide. The method is complementary to the van der Pauw method (see chapter 2) for what concerns the measurement of the sheet resistance and can be useful when it is not possible to reliably place contacts on the sample boundary.

### 1.1.1 Measurement principle

The 4PP method consists in placing four probes evenly spaced in a straight line on the sample to be measured. For ideal conditions no geometrical information is needed to measure $R_S$ [1]. To achieve this the sample should be a 2D infinite layer, i.e. i) the distance between probes, $s$, should be ideally negligible compared to the dimensions of the sample and ii) the distance $x$ between the probes and the sample edge should be large compared to $s$. To retrieve $R_S$, a current $I$ is applied through the two external probes and the voltage $V$ is measured between the internal probe pair. The measured $I$ and $V$ quantities are then used to calculate the sheet resistance using a formula. In practice, this formula may contain corrections, depending on e.g. sample dimensions, inter-probe distance, distance between the probes and the sample edge.

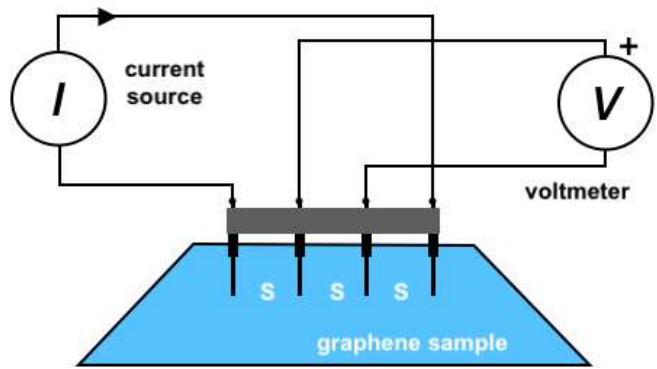

**Figure 1.1 – Schematic representation of the four-point probe method**

### 1.1.2 Sample preparation method

The sample is measured as it is delivered by the supplier. No special sample preparation is required.

The sample has to be as large as possible in order to accomplish to condition of the method of space between probes very small compared to dimensions of the sample. For this, measurements on entire wafers are preferred respect to measurements on individual dice. Rule of thumb: a finite sample can be considered as infinite when the overall width is at least one order of magnitude larger than the probe spacing.

The substrate supporting the graphene layer should be planar and insulating to prevent any contribution of the volume resistivity of the substrate.

### 1.1.3 Description of measurement equipment / apparatus

The equipment necessary to carry out measurements with the in line four probes method consist of a current source and a voltmeter. Alternatively, both instruments can be replaced by an ohmmeter with capability of measuring resistance by the 4-terminal method. The source/measurement range of the instrumentation, for measuring monolayer CVD graphene, should be of 1 μA to 1 mA for the current source and 1 mV to 1 V for the voltmeter, considering typical 5- or 6-digit commercial instruments.

Furthermore, the system must include a fixture to properly hold the sample and place the probes in position.

### 1.1.4 Ambient conditions during measurement

Being a carbon monolayer, graphene electrical properties are strongly affected by ambient conditions, in particular, air humidity [2]. The ambient conditions during the whole duration of the measurement must be monitored and recorded. Typical ambient conditions are those of electrical calibration laboratories, T = (23 ± 1) °C, RH = (50 ± 4) % but can be chosen from the table 1 in of IEC 60068-1. An air bath may be used if it is possible to introduce the sample or the system into it.

## 1.2  Measurement procedure

A schematic diagram of the measurement arrangement is presented in the Fig. 1.1.

### 1.2.1 Calibration of measurement equipment

The current source and the voltmeter must be calibrated and within the specified calibration period.

### 1.2.2 Detailed protocol of the measurement procedure

Turn on the equipment before using it the time indicated in manuals. Instrumentation must be let warm up according to specifications.

Place the four measuring probes aligned on the graphene sample, according to Fig. 1.1. The separation distance between the probes, *s,* must be known, as well as the one between the current probes and the edge, *x*. In this way, check if the probes are equidistant and if their position meet the condition $x \geq 4s$ If so, the simplest case calculations apply. Otherwise correction factors must be calculated. See [3] and [4, Sec. 4] and the following discussion.

The current to be applied during the 4PP resistance measurement must have a value such as it cannot damage the sample, results in value that the voltmeter can measure accurately. Measurement accuracy is improved by performing two readings of each measurement of $V$ with two currents equal in nominal magnitude and opposite in sign, and by averaging the results; this help to reduce thermoelectric effects which introduce systematic errors.

Repeated measurement of V will allow to estimate the measurement standard deviation. Rule of thumb: a reasonable number of repeated measurements should allow to keep the measurement standard deviation below 0.1% for this kind of measurement with commercial laboratory grade instruments.

## 1.2.3 Measurement accuracy

For the definition of uncertainty and related terms used in the following see Ref. [5, Sec. 2.3].

The 4PP measurement accuracy depends on:

- The measurement accuracy, resulting by the combination of the measurement's standard deviation (type-A uncertainty) and instruments accuracy specifications (type-B uncertainty)
- The ambient conditions knowledge and stability.
- Other knowledge about the measurement setup configuration, e.g. probes distance and placement.

**Table 2.1: Example of measurable values for $R_s$, and the corresponding measurement settings and type-B uncertainty, when using a current source Keithley 2602B and a voltmeter HP 34420.**

**(1y calibration specifications)**

| Resistance $R$ / Ω | I applied / A | I range / µA | V meas / V | V range / mV | Relative type-B uncertainty of $R_s$ |
|---|---|---|---|---|---|
| 1E-02 | 1.00E-04 | 100 | 1E-06 | 1 | 1.19E-02 |
| 1E-01 | 1.00E-05 | 10 | 1E-06 | 1 | 9.79E-03 |
| 1E+00 | 1.00E-05 | 10 | 1E-05 | 1 | 1.43E-03 |
| 1E+01 | 1.00E-05 | 10 | 1E-04 | 1 | 5.93E-04 |
| 1E+02 | 1.00E-05 | 10 | 1E-03 | 10 | 5.14E-04 |
| 1E+03 | 1.00E-05 | 10 | 1E-02 | 100 | 5.18E-04 |
| 1E+04 | 1.00E-05 | 10 | 1E-01 | 1000 | 5.18E-04 |
| 1E+05 | 1.00E-06 | 1 | 1E-01 | 1000 | 9.18E-04 |

## 1.2.4 Data analysis / interpretation of results

### Equidistant probes

In the ideal case, the equations for $R_S$ and its uncertainty are as follows:

$$R_S = \frac{V}{I} 4.5324 \ldots \tag{1.1}$$

$$u_{R_S} = \sqrt{u_A^2 + \sum u_B^2} \tag{1.2}$$

where:

$u_A$: is the type-A uncertainty, estimated as the measurement standard deviation.

$u_B$: is the type-B uncertainty, estimated from the instrumentation specifications.

$$u_{R_S} = \sqrt{u_A^2 + \sum u_B^2} = \sqrt{u_m^2 + u_s^2 + u_V^2} \tag{1.3}$$

where:

$u_m$: Uncertainty due to the spread of the voltage measurements (type-A).

$u_I$: Uncertainty due to the specifications of the current source (type-B).

$u_V$: Uncertainty due to the specifications of the voltmeter (type-B).

### Non-equidistant probes

In case of non-equidistant probes, the expression for $R_S$ is as follows (adapted from [3, eq. 2.5])

$$R_S = \frac{V}{I}\left[\frac{2\pi}{\frac{1}{s_1}+\frac{1}{s_3}-\frac{1}{s_1+s_2}-\frac{1}{s_2+s_3}}\right] \tag{1.4}$$

The uncertainty propagation requests to consider the equation (1.4) and implies implicit derivative functions:

$$u^2(R_S) = \left(\frac{\partial R_S}{\partial V}\right)^2 u^2(V) + \left(\frac{\partial R_S}{\partial I}\right)^2 u^2(I) + \left(\frac{\partial R_S}{\partial s_1}\right)^2 u^2(s_1) + \left(\frac{\partial R_S}{\partial s_2}\right)^2 u^2(s_2) + \left(\frac{\partial R_S}{\partial s_3}\right)^2 u^2(s_3)$$

$$\tag{1.5}$$

Calculating the uncertainty in relative mode and operating:

$$\frac{u^2(R_S)}{R_S^2} = \left[\frac{2\pi}{\frac{1}{s_1} + \frac{1}{s_3} - \frac{1}{s_1 + s_2} - \frac{1}{s_2 + s_3}}\right]^2 \frac{u^2(V)}{V^2} + \left[\frac{2\pi}{\frac{1}{s_1} + \frac{1}{s_3} - \frac{1}{s_1 + s_2} - \frac{1}{s_2 + s_3}}\right]^2 \frac{u^2(I)}{I^2}$$

$$+ \left[\frac{2\pi\left(\frac{1}{(s_1 + s_2)^2} - \frac{1}{s_1^2}\right)}{\left(\frac{1}{s_1} + \frac{1}{s_3} - \frac{1}{s_1 + s_2} - \frac{1}{s_2 + s_3}\right)^2}\right]^2 \frac{u^2(s_1)}{s_1^2}$$

$$+ \left[\frac{2\pi\left(\frac{1}{(s_1 + s_2)^2} + \frac{1}{(s_2 + s_3)^2}\right)}{\left(\frac{1}{s_1} + \frac{1}{s_3} - \frac{1}{s_1 + s_2} - \frac{1}{s_2 + s_3}\right)^2}\right]^2 \frac{u^2(s_2)}{s_2^2}$$

$$+ \left[\frac{2\pi\left(\frac{1}{(s_2 + s_3)^2} - \frac{1}{s_3^2}\right)}{\left(\frac{1}{s_1} + \frac{1}{s_3} - \frac{1}{s_1 + s_2} - \frac{1}{s_2 + s_3}\right)^2}\right]^2 \frac{u^2(s_3)}{s_3^2}$$

<div align="right">(1.6)</div>

Where $s_1$, $s_2$ and $s_3$ are the distances between the probes (see figure 1.1). They depend on the specifications and calibration of the length meter.

## Proximity to the edge

In case of proximity to the edge, but with equidistant probes, the expression for $R_S$, adapted from [6, eq. 6], is as follows:

$$R_S = \frac{V}{I}\left[\frac{2\pi}{\frac{1}{s} - \frac{2}{\sqrt{(2s)^2 + (2x)^2}} + \frac{2}{\sqrt{s^2 + (2x)^2}}}\right] \tag{1.7}$$

Where:

$s$ = distance between the probes.

$x$ = distance between the current probes and the edge.

Derivative functions are requested, too:

$$u^2(R_S) = \left(\frac{\partial R_S}{\partial V}\right)^2 u^2(V) + \left(\frac{\partial R_S}{\partial I}\right)^2 u^2(I) + \left(\frac{\partial R_S}{\partial s}\right)^2 u^2(s) + \left(\frac{\partial R_S}{\partial x}\right)^2 u^2(x) \tag{1.8}$$

Calculating the uncertainty in relative mode and operating:

$$\frac{u^2(R_S)}{R_S^2} = \left[\frac{2\pi}{\dfrac{1}{s} - \dfrac{2}{\sqrt{(2s)^2 + (2x)^2}} + \dfrac{2}{\sqrt{s^2 + (2x)^2}}}\right]^2 \frac{u^2(V)}{V^2}$$

$$+ \left[\frac{2\pi}{\dfrac{1}{s} - \dfrac{2}{\sqrt{(2s)^2 + (2x)^2}} + \dfrac{2}{\sqrt{s^2 + (2x)^2}}}\right]^2 \frac{u^2(I)}{I^2}$$

$$+ \left[\frac{2\pi\left(\dfrac{1}{s^2} - \dfrac{8s}{\left(\sqrt{(2s)^2 + (2x)^2}\right)^3} + \dfrac{2s}{\left(\sqrt{s^2 + (2x)^2}\right)^3}\right)}{\left(\dfrac{1}{s} - \dfrac{2}{\sqrt{(2s)^2 + (2x)^2}} + \dfrac{2}{\sqrt{s^2 + (2x)^2}}\right)^2}\right]^2 \frac{u^2(s)}{s^2}$$

$$+ \left[\frac{2\pi\left(\dfrac{8x}{\left(\sqrt{(2s)^2 + (2x)^2}\right)^3} + \dfrac{8x}{\left(\sqrt{s^2 + (2x)^2}\right)^3}\right)}{\left(\dfrac{1}{s} - \dfrac{2}{\sqrt{(2s)^2 + (2x)^2}} + \dfrac{2}{\sqrt{s^2 + (2x)^2}}\right)^2}\right]^2 \frac{u^2(x)}{x^2}$$

(1.9)

Both $u(s)$ and $u(x)$ depend on the specifications and calibration of the used measurement instrument.

### Non-equidistant probes and proximity to the edge

The case in which both conditions are combined is even more complicated. See the bibliography for more specific references.

It is not worthwhile to follow this method in this case. It is better, if possible, to rely on the van der Pauw method (see chapter 2).

## 1.3   Bibliography

[1]   Smits, F. M. "Measurement of sheet resistivities with the four-point probe," Bell System Technical Journal, vol. 37, no. 3, pp 711-718, 1958.

[2]   C. Melios, C. E. Giusca, V. Panchal, and O. Kazakova, "Water on graphene: Review of recent progress," 2D Materials, vol. 5, no. 2, 2018.

[3]     Miccoli, I., Edler, F., Pfnür, H., Tegenkamp, C., "The 100th anniversary of the four-point probe technique: the role of probe geometries in isotropic and anisotropic systems," Journal of Physics: Condensed Matter, vol. 27, no. 22, pp. 223201, 2015.

[4]     Yamashita, M., Agu, M., "Geometrical correction factor for semiconductor resistivity measurements by four-point probe method," Japanese journal of applied physics, vol. 23(11R), pp. 1499, 1984.

[5]     JCGM100:2008, GUM1995, Joint Committee for Guides in Metrology, JCGM100:2008, and GUM1995, "Evaluation of measurement data — Guide to the expression of uncertainty in measurement," JCGM 1002008 GUM 1995 with Minor Correct., 2008.

[6]     Andrew P. Schuetze et al. "A laboratory on the four-point probe technique". Am. J. Physics, vol. 72, no. 2, pp. 149-153, 2004.

# 2 Measurement of sheet resistance of graphene films with the van der Pauw method

## 2.1 General

This part of the Good Practice Guide establishes a method to determine the key control characteristics sheet resistance $R_S$ (measured in $\Omega$/sq), by the van der Pauw method (van der Pauw, 1958). The van der Pauw (vdP) method is a technique that can be applied to a sample of arbitrary shape.

The sheet resistance, carrier mobility and density are derived by measurements of four-terminal electrical resistance performed on four electrical contacts placed on the boundary of the planar sample. During the resistance measurements, the sample can be in a uniform dc magnetic field perpendicular to the sample surface.

The sheet resistance can be derived with a mathematical expression involving two resistance measurements.

The method assumes a uniform and isotropic carrier mobility and carrier density of the sample.

The measurement range for the sheet resistance of the graphene layer shall be in the nominal range of $10^{-2} - 10^{4}$ $\Omega$/sq.

The method is applicable for CVD graphene provided it is transferred to quartz substrates or other rigid, insulating materials as well as graphene on silicon carbide. The method is complementary to the in-line four-contact method for what concerns the measurement of the sheet resistance and can be applied when it is possible to reliably place contacts on the sample boundary.

### 2.1.1 Measurement principle

The measurement principle of the vdP method applied to the assessment of the sheet resistance $R_S$ of graphene films is based on the measurement of two transresistances, performed on four contacts on the sample boundary. A mathematical theorem, derived from conformal mapping theory, relates the transresistance of the sample measured in two different configurations with its sheet resistance [1]. The problem is practically solved with an explicit formula that has as input the two measured four-terminal resistances and as output $R_S$. The method is not sensitive to the sample shape and contacts positioning in principle; corrections for real cases exist.

### 2.1.2 Sample preparation

The sample is measured as it is delivered by the supplier. No special sample preparation is required. An insulating, planar, material must support the graphene; hence graphene synthesized on conductive material must be transferred on an insulating support.

### 2.1.3 Description of measurement equipment / apparatus

The measurement setup for the measurement of $R_S$ with the vdP method consists of:

– a source of dc current;
– a voltmeter;
– a fixture/sample holder;
– an optional switching system for measurement automation;
– an optional acquisition and data processing system.

A block schematic diagram of the typical vdP system is shown in Figure 2.1. The switch connects the current source to one pair of contacts, and the voltmeter to the other pair.

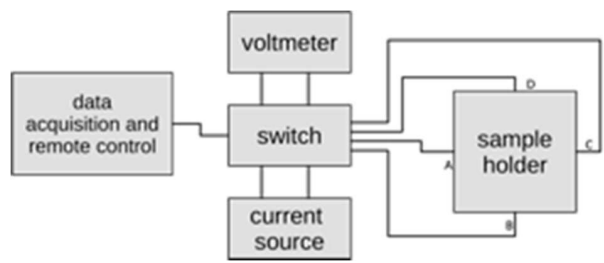

**Figure 2.1: Scheme of a van der Pauw measurement setup.**

The current source must be able to supply current in the range 1 µA to 1 mA.

The voltmeter must be able to measure voltage in the range 1 mV to 1 V.

The sample holder fixture must have purely mechanical contacts. This is in order to avoid additional lithographic steps that inevitably would contaminate the sample surface. Contacts probe must be adjustable in position and height. It is recommended to test symmetric samples and place the electrodes in symmetric positions.

The switch must be a 4 × 4 matrix able to connect each terminal of the source and meter to each of the contacts on the sample.

### 2.1.4 Ambient conditions during measurement

Being a carbon monolayer, graphene electrical properties are strongly affected by ambient conditions, in particular air humidity [2]. The ambient conditions during the whole duration of the measurement must be monitored and recorded. Typical ambient conditions are those of electrical calibration laboratories, T = (23 ± 1) °C, RH = (50 ± 4) % but can be chosen from the table 1 in of IEC 60068-1.

## 2.2    Measurement procedure

### 2.2.1  Calibration of measurement equipment

The current source and the voltmeter must be calibrated and within the specified calibration period.

### 2.2.2  Detailed protocol of the measurement procedure

Quantities definition:

$I_{ij}$ : current applied to a pair of contacts entering contact i, and exiting at contact j.

$V_{kl}$ : voltage measured between contacts k and l.

$R_{ij,kl}$ : four-terminal resistance calculated as  $I_{ij}$ / $V_{kl}$.

$R_S$ : van der Pauw sheet resistance calculated from $R_{ij,kl}$.

In the following we call the four contacts A,B,C,D. The contacts must be placed on the sample in counter clock-wise order along the sample boundary (see labels in figure 2.2).

The four-terminal resistance measurements to be performed are

$$R_{\mathrm{AB,CD}} = \frac{V_{\mathrm{CD}}}{I_{\mathrm{AB}}} \tag{2.1}$$

$$R_{\mathrm{BC,DA}} = \frac{V_{\mathrm{DA}}}{I_{\mathrm{BC}}} \tag{2.2}$$

These are measured following the configurations reported in figure 2.2-a and –b respectively.

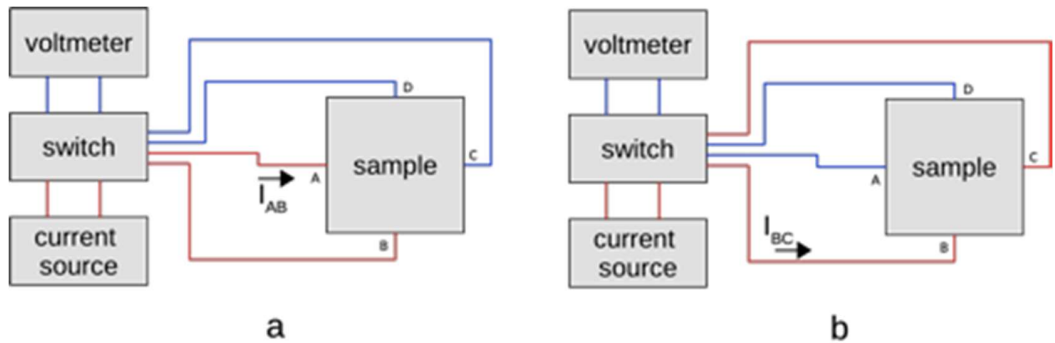

**Figure 2.2: Schematic of a typical vdP measurement setup in the two measurement configurations,** $R_{\mathrm{AB,CD}}$ **(a) and** $R_{\mathrm{BC,DA}}$ **(b). Red leads carry the current** $I_{ij}$**, blue ones are used for the measurement of the corresponding** $V_{kl}$**.**

**Settings and precautions for the measurement of** $R_{ij,kl}$

- The magnitude of $I_{ij}$ is a trade-off between the need to achieve a $V_{kl}$ magnitude that can be measured with enough accuracy by the voltmeter, and the need to avoid sample heating on the contact point. Heating effects can be identified, by performing measurements at two different current levels (e.g., *I* and *2I*) and comparing the measurement outcome. High contact resistance may produce sample heating even at low injecting currents. Heating of the sample exposed to the laboratory ambient may also cause sample oxidation which could alter irreversibly the sample's properties [3,4].

— A too large magnitude of $I_{ij}$ can also cause an abrupt breakdown of the sample. It has been reported that, for CVD graphene samples having sheet resistance in the range of 1000 $\Omega$/sq the breakdown current is in the range of $10^7$ A /cm$^2$ [3,4]. A 100 $\mu$A current injected through 50 $\mu$m diameter needle probe would produce a current density of $4 \times 10^4$ A/cm$^2$, well below the reported breakdown level.

— The current source must be set in manual range mode. The current source must be switched on with zero current level before connecting the contacts to the sample to avoid any voltage transient that could damage the sample.

— Measurement accuracy is improved by performing two readings of $R_{ij,kl}$ with two currents equal in nominal magnitude and opposite in sign, and by averaging the results.

— Measurement accuracy is improved, and type A contribution to measurement uncertainty estimated, by recording time series of individual measurements and computing averages and standard deviations.

Note: for a symmetric sample in principle $R_{AB,CD} = R_{BC,DA}$ hence only a single measurement would be necessary (e.g. $R_{AB,CD}$). In practice samples are never perfectly symmetric in shape and concerning the position of the contacts; as a consequence, a two-measurement protocol is suggested.

### 2.2.3  Measurement accuracy

The accuracy of the sheet resistance measured with the vdP method depends on the uncertainty of the two transresistance measurements. For the definition of uncertainty and related terms used in the following see Ref. [5, sec. 2.3].

The combined uncertainty $u_{R_S}$ is obtained from the type-A and type-B uncertainty components of the measurement of $R_{ij,kl}$.

$$u_{R_{ij,kl}} = \sqrt{\left(u_{A,R_{ij,kl}}\right)^2 + \left(u_{B,R_{ij,kl}}\right)^2} \tag{2.3}$$

Type-A uncertainty $u_{A,Rij,kl}$, or standard uncertainty, is in this case the standard deviation of the repeated series measurement. The type-B uncertainty component $u_{B,Rij,kl}$, is calculated from the instruments' documentation and will depend on the absolute values of the measured transresistance $R_{ij,kl}$. In table 1 are reported possible transresistance values and the current and voltage range needed to measure them with standard instrumentation. CVD graphene typically shows a sheet resistance within the range of 10 - 100 $\Omega$ (shaded rows in table 2.1).

**Table 2.1: Example of measurable transresistance values for $R_{ij,kl}$, and the corresponding measurement settings and type B uncertainty, using a current source Keithley 2602B and a voltmeter HP 34420 (within 1y calibration specifications)**

| Resistance $R$ / Ω | I applied / A | I range / µA | V meas / V | V range / mV | Relative uncertainty of $R$ |
|---|---|---|---|---|---|
| 1E-02 | 1.00E-04 | 100 | 1E-06 | 1 | 1.19E-02 |
| 1E-01 | 1.00E-05 | 10 | 1E-06 | 1 | 9.79E-03 |
| 1E+00 | 1.00E-05 | 10 | 1E-05 | 1 | 1.43E-03 |
| 1E+01 | 1.00E-05 | 10 | 1E-04 | 1 | 5.93E-04 |
| 1E+02 | 1.00E-05 | 10 | 1E-03 | 10 | 5.14E-04 |
| 1E+03 | 1.00E-05 | 10 | 1E-02 | 100 | 5.18E-04 |
| 1E+04 | 1.00E-05 | 10 | 1E-01 | 1000 | 5.18E-04 |
| 1E+05 | 1.00E-06 | 1 | 1E-01 | 1000 | 9.18E-04 |

## 2.2.4  Data analysis / interpretation of results

### Calculation of $R_S$

The sheet resistance $R_S$ is calculated from $R_{AB,CD}$ and $R_{BC,DA}$ by means of the formula

$$R_S = \frac{\pi}{ln2} \frac{R_{AB,CD} + R_{BC,DA}}{2} f \tag{2.4}$$

Where $f$ is a numerical factor, function of the ratio $R_{AB,CD} / R_{BC,DA}$ , defined in ref. [1], and reported in table in ref [6].

For the case in which the sample geometry is symmetric, it is expected that $R_{AB,CD} / R_{BC,DA} = 1$ ($f = 1$), and the equation (2.4) simplifies as

$$R_S = \frac{\pi}{ln2} \frac{R_{AB,CD} + R_{BC,DA}}{2} \tag{2.5}$$

### Further corrections

Depending on the contacts position (distance from physical edge) and size, further corrections may be applied [7, 8, 9].

For square samples with contacts at the four corners (symmetric samples), in particular:

– If contacts are placed within a distance d from the border for which the ratio $d/D$ is 0.05, where $D$ is the lateral size of the sample,
– The size of the contact is such that the ratio β/$D$ < 0.1, where β is the contact lateral size and $D$ is the lateral size of the sample, no further correction must be calculated, because their entity becomes negligible compared to the instrumental accuracy given in table 2.1. The above conditions are very easy to meet with needle probe contacts.

## Expression of uncertainty on $R_S$

The uncertainty $u_{R_S}$ of $R_S$ depends on the measurement accuracy of $R_{AB,CD}$ and $R_{BC,DA}$.

In the general case of a non-symmetric samples with $R_{AB,CD} / R_{BC,DA} \neq 1$ ($f > 1$), the uncertainty shall be calculated following the general formula for the propagation of uncertainty [5]

$$u_{g(x_i)} = \sqrt{\sum_i \left( \frac{\partial g(x_i)}{\partial x_i} u_{x_i} \right)^2} \qquad (2.6)$$

where $g(x_i)$ is an arbitrary function of several variables $x_i$. In the case of vdP method eq. (2.6) reads

$$u_{R_S} = \frac{\pi}{2 ln 2} \sqrt{(f \cdot u_{R_{BC,DA}})^2 + (f \cdot u_{R_{AB,CD}})^2 + \left( \frac{\partial f}{\partial R_{AB,CD}} \cdot u_{R_{AB,CD}} \right)^2 + \left( \frac{\partial f}{\partial R_{BC,DA}} \cdot u_{R_{BC,DA}} \right)^2} \qquad (2.7)$$

which contains partial derivatives of $f$ which is an implicit function of $R_{AB,CD}$ and $R_{BC,DA}$ [1].

In this case two scenarios are possible:

1) $R_{AB,CD} / R_{BC,DA} \sim 1$, $f \sim 1$ and the partial derivatives of $f$ are negligible.
2) $R_{AB,CD} / R_{BC,DA} \neq 1$, the derivative of $f$ is not negligible.

For the case 1) consider that if $R_{AB,CD}$ and $R_{BC,DA}$ and are equal within 1%, $f$ is equal to 1 within 0.001% and can be neglected [10], $R_S$ can be calculated with eq. (2.5). Hence eq. (2.7) get simplified and the calculation of the uncertainty $u_{R_S}$ is

$$u_{R_S} = \frac{\pi}{2 ln 2} \sqrt{\left( u_{R_{BC,DA}} \right)^2 + \left( u_{R_{AB,CD}} \right)^2} \qquad (2.8)$$

For the case 2) the uncertainty propagation requests to consider the full vdP form eq. (2.4) and implies implicit derivative functions calculations of eq. (2.7) not presented here.

## 2.3    Bibliography

[1]    L. J. van der Pauw, "Method of measuring resistivity and Hall coefficient on lamellae of arbitrary shape," Philips Tech. Rev., vol. 20, no. 8, pp. 220–224, 1959.

[2]    C. Melios, C. E. Giusca, V. Panchal, and O. Kazakova, "Water on graphene: Review of recent progress," 2D Materials, vol. 5, no. 2, 2018.

[3]     X. Chen et al., "Breakdown current density of CVD-grown multilayer graphene interconnects," IEEE Electr. Device L., vol. 32, no. 4, pp. 557-559, 2012.

[4]     X. Chen, D. H. Seo, S. Seo, H. Chung, and H. S. P. Wong, "Graphene interconnect lifetime: a reliability analysis," IEEE Electr. Device L., vol. 33, no. 11, pp. 1604 - 1606, 2012.

[5]     JCGM100:2008, GUM1995, Joint Committee for Guides in Metrology, JCGM100:2008, and GUM1995, "Evaluation of measurement data — Guide to the expression of uncertainty in measurement," JCGM 1002008 GUM 1995 with Minor Correct., 2008.

[6]     A. A. Ramadan, R. D. Gould, and A. Ashour, "On the Van der Pauw method of resistivity measurements," Thin Solid Films, vol. 239, no. 2, pp 272 - 275, 1994.

[7]     D. S. Perloff, "Four-point sheet resistance correction factors for thin rectangular samples," Solid-State Electron., Elsevier, vol. 20, pp. 681-687, 1977.

[8]     W. Versnel, "Analysis of symmetrical Van der Pauw structures with finite contacts," Solid-State Electron., vol. 21, pp. 1261-1268, 1978

[9]     R. Chwang, B. Smith, and C. Crowell, "Contact size effects on the van der Pauw method for resistivity and hall coefficient measurement," Solid-State Electron., vol. 17, no. 12, pp. 1217–1227, 1974.

[10]    G. Rietveld, et al., "DC conductivity measurements in the van der Pauw geometry," IEEE Trans. Instrum. Meas., vol. 52, no. 2, pp. 449–453, 2003.

# 3 Measurement of charge carrier mobility of graphene films with the van der Pauw method with a permanent magnet

## 3.1 General

This part of the Good Practice Guide establishes a method to determine the carrier mobility, $\mu$, of a graphene sample of arbitrary shape.

### 3.1.1 Measurement principle

The measurement principle of the vdP method [1] applied to the assessment of $\mu$ of graphene films is based on the measurement of two four-terminal resistances, performed on four contacts on the sample boundary, measured without and with applied magnetic field. The algebraic expression that returns $\mu$ has as input quantities two measured four-terminal resistances, $R_S$ measured as described in chapter 2, and the intensity of the applied magnetic field $B$. The method is not sensitive to the sample shape and contacts positioning in principle; corrections for real cases exist.

### 3.1.2. Sample preparation

Same as vdP for the measurement of $R_S$.

### 3.1.3. Description of measurement equipment / apparatus

The measurement setup for the measurement of $\mu$ with the vdP method consists of:

- – a source of dc;
- – a nano-voltmeter
- – a permanent magnet;
- – A Hall probe sensor;
- – a fixture/sample holder;
- – an optional switching system for measurement automation;
- – an optional acquisition and data processing system.

A block schematic diagram of the typical vdP system is shown in Figure 3.1. The switch connects the current source to one pair of contacts, and the voltmeter to the other pair. A seat for a permanent magnet is located within the sample holder.

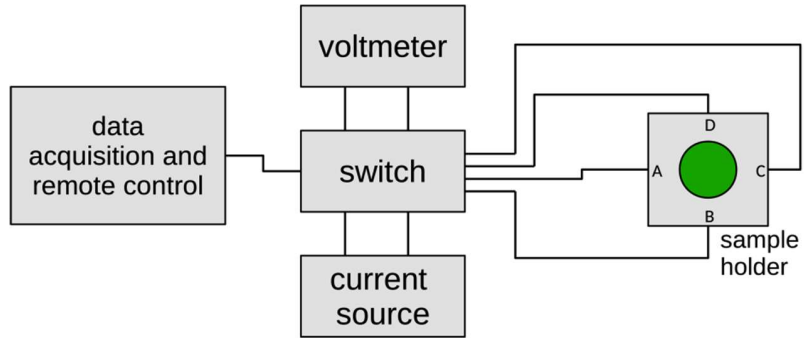

**Figure 3.1: Scheme of a van der Pauw measurement setup for the measurement of the mobility; the sample holder (square) must be able to host a permanent magnet (disc).**

The current source must be able to supply current in the range from 1 µA to 1 mA.

The voltmeter must be able to measure voltage in the range from 1 nV to 1 V.

The magnet must be able to produce a magnetic field in the range of 100-1000 mT.

The sample holder fixture must have purely mechanical contacts to avoid additional lithographic steps that inevitably would contaminate the sample surface. Each contact probe must be adjustable in position and height. It is recommended to test symmetric samples and place the electrodes in symmetric positions.

The switch must be a 4 x 4 matrix able to connect each terminal of the source and meter to each of the contacts on the sample.

### 3.1.4  Ambient conditions during measurement

Same as vdP for the measurement of $R_s$.

## 3.2    Measurement procedure

### 3.2.1  Calibration of measurement equipment

The current source and the voltmeter must be calibrated and within the specified calibration period.

### 3.2.2. Detailed protocol of the measurement procedure

Quantities definition:

- $I_{ij}$ : current applied to a pair of contacts entering contact i, and exiting at contact j.
- $V_{kl}$ : voltage measured between contacts k and l .
- $R^b_{ij,kl}$ : four-terminal resistance calculated as $I_{ij}/V_{kl}$ with or without applied magnet $b$ = 0,1.
- $\mu$ : van der Pauw mobility calculated from { $R^b_{ij,kl}$ }.
- $B$: the magnetic field at sample position.

In the following we call the four contacts {A,B,C,D}. The contacts must be placed on the sample in counter clock-wise order along the sample boundary (see labels in figure 3.2).

The four-terminal resistance measurements to be performed are:

$$R^0_{\mathrm{AC,BD}} = \frac{V_{\mathrm{BD}}}{I_{\mathrm{AC}}} \;\; (B\ \mathrm{off}) \tag{3.1a}$$

$$R^1_{\mathrm{AC,BD}} = \frac{V_{\mathrm{BD}}}{I_{\mathrm{AC}}} \;\; (B\ \mathrm{on}) \tag{3.1b}$$

These are respectively measured as described in Figure 3.2-a and -b.

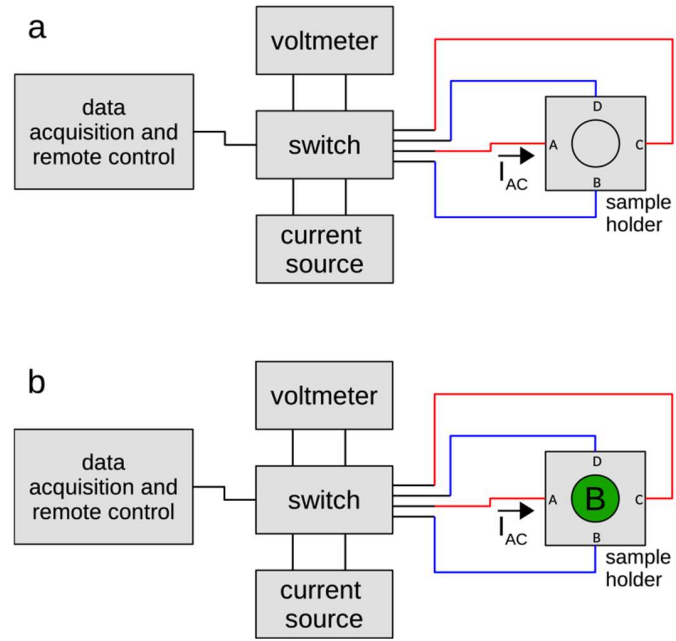

**Figure 3.2: Measurement configuration without (a) and with (b) the permanent magnet in place.**

Initially, the sample transresistance $R^0_{\mathrm{ACBD}}$ is measured without magnetic field by applying a bias current $I_{\mathrm{AC}}$ and measuring the voltage $V_{\mathrm{BD}}$. Then the Hall effect is induced by placing a permanent magnet (green disc) that produces a magnetic field $B$. The sample transresistance $R^1_{\mathrm{ACBD}}$ is measured with the magnetic field on. The magnet seat should be accessible without moving the sample.

The intensity of the magnetic field at the sample position must be measured accurately in order not to introduce systematic errors larger than the uncertainty contribution coming from the electrical measurements. A Hall probe can be used, which should guarantee enough accuracy. The magnetic field intensity should be measured before mounting the sample, by placing the Hall probe sensor in place of the sample itself.

**Settings and precautions to be adopted for the measurement of $R_{ij,kl}$:**

— The magnitude of $I_{ij}$ is a trade-off between the need to achieve a $V_{kl}$ magnitude that can be measured with enough accuracy by the voltmeter, and the need to avoid sample heating on the contact point. Heating effects can be identified, by performing measurements at two different current levels (e.g., $I$ and $2I$) and comparing the measurement outcome. High contact resistance may produce sample heating even at low injecting currents. Heating of the sample exposed to the laboratory ambient may also cause sample oxidation which could alter irreversibly the sample's properties [2.3].

— A too large magnitude of $I_{ij}$ can also cause an abrupt breakdown of the sample. It has been reported that, for CVD graphene samples having sheet resistance in the range of 1000 $\Omega$/sq the breakdown current is in the range of $10^7$ A /cm$^2$ [2.3]. A 100 µA current injected through 50µm diameter needle probe would produce a current density of $4 \times 10^4$ A/cm$^2$, well below the reported breakdown level.

— The current source must be set in manual range mode. The current source must be switched on with zero current level before connecting the contacts to the sample to avoid any voltage transient that could damage the sample.

— Measurement accuracy is improved by performing two readings of $R^b_{ij,kl}$ with two currents equal in nominal magnitude and opposite in sign, and by averaging the results.

— Measurement accuracy is improved, and type-A contribution to measurement uncertainty estimated, by recording time series of individual measurements and computing averages and standard deviations.

If measurement noise limits the measurement of the Hall voltage, some variant of the measurement can be implemented:

— In some case the transverse voltage drop $V_{kl}$ is too weak to be measured with acceptable stability in dc regime with a nano-voltmeter without increasing the bias current $I_{ij}$ too much. Then $V_{kl}$ can be measured in ac regime with an AC current generator and a phase sensitive detection approach, using a lock-in amplifier to measure $V_{kl}$. The Lock-in reference channel is fed with a reference signal at the same frequency of the current source. Note that the overall accuracy of a lock-in is far worse than the accuracy of a dc nano-voltmeter on the other hand. This is recommended only in case of measurement noise is a limit to the measurement itself.

— Another solution to deal with noisy measurements, is to keep a dc current source, and using an electromagnetic coil to generate an AC magnetic field. The lock-in amplifier used to measure the Hall voltage is referenced to a signal of the frequency of the modulation of the magnetic field. Figure 3.3 shows a schematic diagram of the implementation. As in Figure 3.2, measurements are performed with and without applied magnetic field.

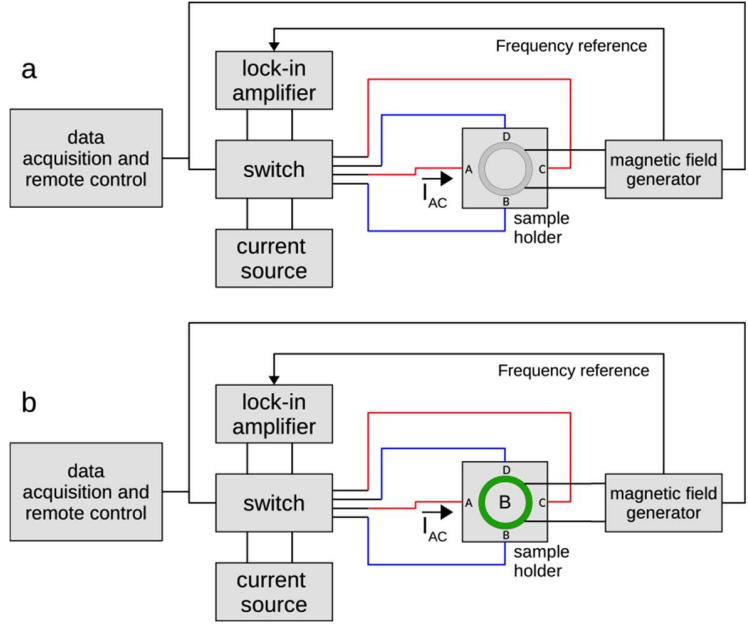

**Figure 3.3: Measurement configuration without (a) and with (b) applied magnetic field.**

— If working with an AC current, a frequency dependence test on the transverse voltage $V_{kl}$ must be done to avoid possible errors.

Note: the same precautions adopted in the static measurement should be taken in account anyway.

### 3.2.3 Measurement accuracy

The accuracy of the mobility measured with the vdP method depends on 1) the accuracy of $R_S$, 2) the accuracy on the Hall resistance measurements $R^b_{ij,kl}$ and 3) the accuracy of the measurement of the magnetic field $B$.

For the definition of uncertainty and related terms used in the following see Ref. [4, Sec. 2.3].

The uncertainty on $R_S$ is expressed in section 2.2.3.

The combined uncertainty of $R^b_{ij,kl}$ is obtained from the type-A and type-B uncertainty components of the measurements:

$$u_{R^b_{ij,kl}} = \sqrt{\left(u_{A,R^b_{ij,kl}}\right)^2 + \left(u_{B,R^b_{ij,kl}}\right)^2} \qquad (3.3)$$

Type-A uncertainty $u_{A,R}$, or standard uncertainty, is in this case the standard deviation of the repeated series measurement. The type-B uncertainty component $u_{B,R}$, is calculated from the instruments' documentation and will depend on the absolute values of the measured Hall resistance $R^b_{ij,kl}$. In table 3.1 are reported possible Hall resistance values and the current and voltage range needed to measure them with standard instrumentation. Depending on the magnetic field intensity, CVD graphene typically shows Hall resistance within the range of 0.01 - 1 Ω.

**Table 3.1: Example of measurable Hall resistance values for $R^b_{ij,kl}$ ($R$ in column 1), the corresponding measurement settings and type-B uncertainty due to the electrical measurement instrumentation, when using a current source Keithley 2602B and a voltmeter HP 34420 (within 1y calibration specifications).**

| $R$ / Ω | I applied / A | I range / µA | V meas / V | V range / mV | Relative uncertainty of $R$ |
|---------|---------------|--------------|------------|--------------|------------------------------|
| 1E-02 | 1.00E-05 | 10 | 1E-07 | 1 | 2.05E-01 |
| 1E-01 | 1.00E-05 | 10 | 1E-06 | 1 | 2.10E-02 |
| 1E+00 | 1.00E-05 | 10 | 1E-05 | 1 | 2.55E-03 |
| 1E+01 | 1.00E-05 | 10 | 1E-04 | 1 | 7.05E-04 |
| 1E+02 | 1.00E-05 | 10 | 1E-03 | 1 | 5.20E-04 |
| 1E+03 | 1.00E-05 | 10 | 1E-02 | 10 | 5.03E-04 |

The uncertainty on $B$ depends on the type of Hall probe used to the measurement of the field intensity. Commercial Hall probe sensors have typically sensitivity of 30 mV/T, which means that for permanent magnets of the order of 300 mT, voltage of about 10 mV are measured at the sensor terminals. This means that using the same voltmeter used for the other voltage measurements in this protocol the accuracy on $B$ would be in the order of $10^{-4}$ in relative terms, which means of the same order of $u_{R_S}$.

## 3.2.4 Data analysis / interpretation of results

### Calculation of the carrier mobility $\mu$

The carrier mobility $\mu$ is given by

$$\mu = \frac{\Delta R}{B R_S} \tag{3.4}$$

Where $\Delta R$ is the resistance difference $R^0_{AC,BD} - R^1_{AC,BD}$, due to the Hall effect, $B$ is the intensity of the magnetic field, and $R_S$ is the sheet resistance measured as described in chapter 2.

### Expression of the uncertainty on $\mu$

The uncertainty $u_\mu$ of $\mu$ depends on the measurement accuracy of $R_S$, $\Delta R_S$ and $B$.

Concerning $R_S$, for uncorrelated $R_{AB,CD}$ and $R_{BC,DA}$ that are equal within 1%, $f$ is equal to 1 within 0.001% and can be neglected [10], $R_S$ can be calculated with eq. (2.4). Hence eq. (2.6) gets simplified and the calculation of the uncertainty $u_{R_S}$ is

$$u_{R_S} = \frac{\pi}{2 ln2} \sqrt{\left(u_{R_{BC,DA}}\right)^2 + \left(u_{R_{AB,CD}}\right)^2} \tag{3.5}$$

Note: For the case of correlated transresistances $R_{AB,CD}$ and $R_{BC,DA}$ the uncertainty propagation requests to consider the full vdP form eq. (2.4) and implies implicit derivative functions calculations as in eq. (2.7).

As regards $\Delta R$, the combined uncertainty $u_{\Delta R_s}$ is the sum of the combined uncertainties of $R^0{}_{AC,BD}$ and $R^1{}_{AC,BD}$

$$u_{\Delta R_s} = \sqrt{\left(u_{R^0_{AC,BD}}\right)^2 + \left(u_{R^1_{AC,BD}}\right)^2} \tag{3.6}$$

The uncertainty on $B$, $u_B$, can be expressed as the type B uncertainty of the Hall voltage measurement at the output terminal of the Hall probe.

Hence the combined uncertainty on $\mu$ is

$$u_\mu = \sqrt{u_{R_s}^2 + u_{\Delta R}^2 + u_B^2} \tag{3.7}$$

Looking at table 3.1 chapter 3, it can be seen that $u_{\Delta R_s}$ is likely to be larger than $u_{R_s}$ and $u_B$.

## 3.3   Bibliography

[1]     L. J. van der Pauw, "Method of measuring resistivity and Hall coefficient on lamellae of arbitrary shape," Philips Tech. Rev., vol. 20, no. 8, pp. 220 – 224, 1959.

[2]     C. Melios, C. E. Giusca, V. Panchal, and O. Kazakova, "Water on graphene: Review of recent progress," 2D Materials, vol. 5, no. 2, 2018.

[3]     X. Chen et al., "Breakdown current density of CVD-grown multilayer graphene Interconnects," IEEE Electr. Device L., vol. 32, no. 4, pp. 557-559, 2012.

[4]     JCGM100:2008, GUM1995, Joint Committee for Guides in Metrology, JCGM100:2008, and GUM1995, "Evaluation of measurement data — Guide to the expression of uncertainty in measurement," JCGM 1002008 GUM 1995 with Minor Correct., 2008.

# 4 Electrical Resistance Tomography of graphene layers

## 4.1 General

This part of the Good Practice Guide describes a method to determine the key control characteristic sheet resistance $R_S$ (measured in $\Omega$/sq) using electrical resistance tomography (ERT). This method is applicable for measuring the spatial distribution of the conductivity, $\sigma(x,y)$, expressed in units of siemens S, of graphene films supported on a rigid insulating material and evaluate the sheet resistance $R_S$ as the map average of $\sigma(x,y)$. The technique is non-destructive and does not require lithographic steps [1].

With this method a map of $\sigma(x,y)$ is obtained from a series of four-terminal electrical transresistance measurements made on the sample surface. This method allows direct metrological traceability to the *SI*. The method is applicable for graphene grown by CVD on $Si/SiO_2$ or transferred on any rigid, insulating substrate. The spatial resolution of ERT is limited by the number of electrodes (e.g. for a 16-electrode array on a $1 \times 1$ cm$^2$ sample it is about 400 µm). The map resolution can be increased by placing more electrodes. The measurement range for the sheet resistance $R_S$ of the graphene layer shall be in the nominal range of $10 - 5 \times 10^3$ $\Omega$/sq.

### 4.1.1 Measurement principle

The electrical resistance tomography is based on the solution of an electrostatic problem in which the conductivity distribution of an object is estimated based on boundary transresistance measurements [2,3]. By means of iterative numerical methods, starting from a guess conductivity distribution, the final $\sigma(x,y)$ is reconstructed [3,4]. To do that, the difference between the measured transresistance and calculated values at each step are minimized. The outcome of ERT (measurements + calculations) is a *conductivity map - a color map of $\sigma(x,y)$*, which is also called *reconstructed image* in the following, from the spatial average reciprocal of which the sheet resistance $R_S$ is obtained.

### 4.1.2 Sample preparation method

The sample is measured as it is delivered by the supplier. No special sample preparation is required. An insulating, planar, material must support the graphene; hence graphene synthesized on conductive material must be transferred on an insulating support.

### 4.1.3 Description of measurement equipment / apparatus

The measurement setup for the ERT method consists of:

- a source of dc current;
- a voltmeter;
- a fixture/sample holder;
- a switching system for measurement automation;
- an acquisition and data processing system.

A schematic of an ERT probe configuration and a picture of a probe fixture are respectively shown in Figure 4.1-a and -b. The ERT fixture consists of *n* spring-mounted metal needle probes with uniform tip radius.

Measurements are performed in several 4-contact configurations. The current source (*I*) supplies current through a pair of probes, and a voltmeter (*V*) measures the voltage across another pair of probes to measure the sample transresistance. A relay switch provides the different combinations of 4-terminal measurements. The current source, voltmeter and switch are commercially available, while the sample-holder is custom. The measurements are automated with remote instruments programming [5].

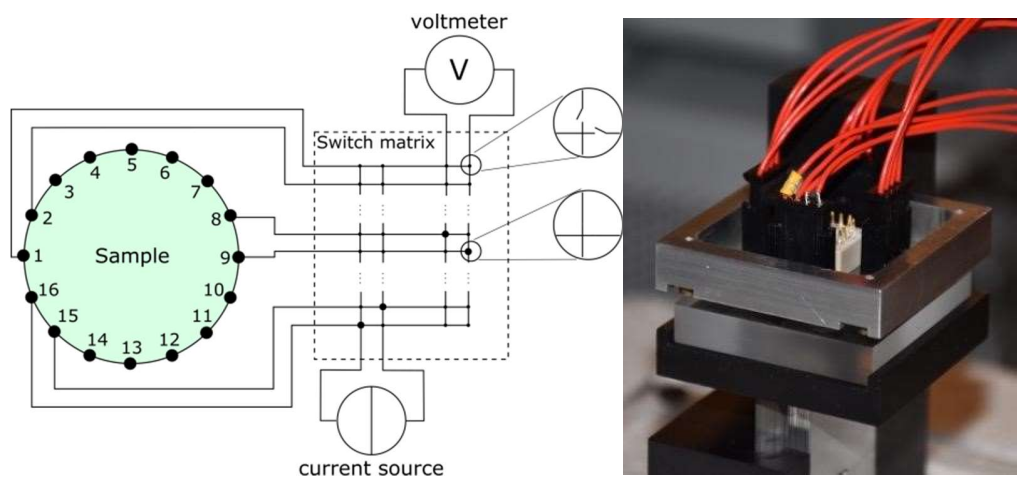

**Figure 4.1 – Scheme of the experimental setup (a) and a picture of the sample-holder with the contact fixture on top (b) sample is seated in the plastic block between the black base and the aluminium head**

The current source must be able to supply current in the range 100 nA to 1 mA with proper accuracy.

The voltmeter must be able to measure voltage in the range 1 µV to 100 mV with proper accuracy.

The sample holder fixture must have purely mechanical contacts. This in order to avoid additional lithographic steps that inevitably would contaminate the sample surface. Contact probes should be adjustable in position and height. Contact-fixture should be engineered in order to place the contacts as near as possible to the sample's boundary. This because even though ERT does not require strictly boundary measurements, the information of the outermost area would be lost in during the image reconstruction. The switch must be a *4 × n* matrix able to connect each terminal of the source and meter to each of the *n* contacts on the sample.

### 4.1.4  Ambient conditions during measurement

Being a carbon monolayer, graphene electrical properties are strongly affected by ambient conditions, in particular the humidity [6].  The ambient conditions during the whole duration of the measurement must be monitored and recorded. Typical ambient conditions are those of electrical calibration laboratories, T = (23 ± 1) °C, RH = (50 ± 4) % but can be chosen from the table 1 in of IEC 60068-1.

## 4.2 Measurement procedure

### 4.2.1 Calibration of measurement equipment

The current source and the voltmeter must be calibrated and within the specified calibration period.

### 4.2.2 Detailed protocol of the measurement procedure

Input quantities for the ERT numerical solvers are a series of transresistance measurements that can be obtained following a so-called stimulation/measurement pattern [7].

To perform the measurements the sample is placed into an engraved seat (e.g. the plastic block in figure 4.1-b). The probe tips contact the sample through adjustment of the height of the stage by a lever. Physical contact between the point probe and sample surface is assured by the probe's springs action with a travel of 1 mm.

Quantities definition:

- $I_{ij}$ : current applied to a pair of contacts entering contact $i$, and exiting at contact $j$.
- $V_{kl}$ : voltage measured between contacts $k$ and $l$.
- $R_{ij,kl}$ : four-terminal resistance calculated as $I_{ij}/V_{kl}$ .

In the following we label the contacts with integers $i,j,k,l$ form 1 to $n$. The contacts are placed on the sample boundary (see labels in figure 4.1-a).

The four-terminal resistance measurements to be performed are of the type

$$R_{ij,kl} = \frac{V_{kl}}{I_{ij}} \tag{4.1}$$

Typically, in ERT, a so-called adjacent/adjacent stimulation/measurement pattern (AD/AD) is followed. In the AD/AD protocol, one pair $\{i,j\}$ of adjacent contacts is used to inject the current $I_{ij}$, while another pair $\{k,l\}$ is used to measure the voltage. Depending on the number of electrodes, the total number of 4-terminal measurements with the AD/AD pattern is $N= n(n-3)$, where n is the number of electrodes. The electrode number is $n = 16$ in the present case so $N = 208$.

In figure 4.1 is represented one of the 208, 4-terminal measurement configurations, corresponding to the measurement $R_{15,16;8,9}$, I.e. current supplied through contacts {15,16} and voltage measured at pair {8,9}.

**Settings and precautions to be adopted for the measurement of $R_{ij,kl}$:**

- The magnitude of $I_{ij}$ is a trade-off between the need to achieve a $V_{kl}$ magnitude that can be measured with enough accuracy by the voltmeter, and the need to avoid sample heating at contacts. Heating effects can be identified, by performing measurements at two different current levels (e.g., $I$ and $2I$) and comparing the measurement outcome. High contact resistance may produce sample heating even at low injecting currents. Heating of the sample exposed to the laboratory ambient may cause sample oxidation which would change the measurement outcome [8,9].

- A too large magnitude of $I_{ij}$ can also cause an abrupt breakdown of the sample. It has been reported that, for CVD graphene samples having sheet resistance in the range of 1000 Ω/sq the breakdown current is in the range of $10^7$ A /cm$^2$ [8,9].  A 100μA current injected through 50 μm diameter needle probe would produce a current density of $4 \times 10^4$ A/cm$^2$, well below the reported breakdown level.
- The current source must be set in manual range mode. The current source must be switched on with 0 A level before connecting the contacts to the sample to avoid any voltage transient that could damage the sample.
- Measurement accuracy is improved by performing two readings of $R_{ij,kl}$ with two currents equal in magnitude and opposite in sign, and by averaging the results.
- Measurement accuracy is improved, and type A contribution to measurement uncertainty estimated, by recording time series of individual measurements and computing averages and standard deviations.

### 4.2.3  Measurement accuracy

ERT is a multi-parametric optimization problem. This makes complex to trace the accuracy of the reconstructed $\sigma(x,y)$ map and hence of $R_S$ to the measured transresistance series. Moreover, regularization techniques tend to smooth the solution and limit the spatial resolution. This implies that sharp changes in resistance (which are likely in graphene) will be always relatively smeared out on the map. This represents an additional problem in the assignment of an accuracy value to each pixel of the map.

Nevertheless, the lower is the measurement noise, the lower is the minimum required regularization and the more reliable is the reconstructed map. Hence accurate electrical measurements should guarantee, provided a correct numerical treatment, the optimal image reconstruction.

At least it is possible to consider the overall transresistance measurements accuracy as an indicator of an accurate reconstruction.

Accuracy on the transresistance measurement can be calculated as for the vdP method. Transresistance measurements with accuracy better than 0.1 % in relative terms are enough to perform ERT without limitations due to the electrical measurements.

### 4.2.4  Data analysis / interpretation of results

ERT image reconstruction can be obtained with several numerical approaches [10,11,12]. A finite element model (FEM) of the sample geometry and contact placement must be generated. The mesh must be dense enough (element size ~ 10% of the interelectrode distance). The model is used to calculate projected transresistances, $R_p$, starting from a guess $\sigma_0(x,y)$ distribution  (e.g. initially uniform). Iteratively the difference between the projected transresistances and the measured ones ($R_m$) is minimized and the distribution $\sigma_i(x,y)$ is updated until a threshold is reached. The problem of guessing $R(x,y)$ from the transresistance data is called ill-posed [2,3]. To deal with ill-posed nature of the problem, it must be regularized, and the optimization problem reads

$$\min_{\sigma} \left\| R_m - R_p(\sigma) \right\|^2 + \lambda \left\| \mathbf{M} \right\|^2 \qquad (4.3)$$

Where λ is the regularization parameter, and the matrix **M** is the regularization term which imposes bounds on the solution and also may contain prior information about the expected $\sigma(x,y)$. With a too small λ the image reconstruction may fail. Increasing the regularization term has the effect of damping the effects of noise in the solution of the inverse problem, but also smooths the reconstructed image. A too large λ yields flat and featureless images.

The choice of the regularization term is then critical for a meaningful reconstruction; many methods exist to search for the optimal regularisation term given a model and a data set. For a review see refs [12, 13].

A typical ERT map of a monolayer $1 \times 1\ cm^2$ CVD graphene sample interpolated to a $100 \times 100$ pixel grid is shown in figure 5.1.

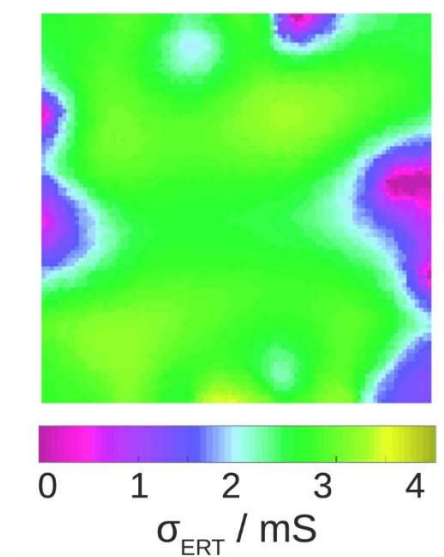

**Figure 4.2 – Conductivity map of a $1 \times 1\ cm^2$ CVD graphene sample reconstructed with ERT. The colour scale represents the conductivity distribution. Adapted from *Sci. Rep.* 9, 10655, 2019 [1].**

The map in figure 5.1 was obtained using the MATLAB routines library EIDORS [4] that includes all the functions to perform the following steps and obtain a sheet resistance map are:

3) Make a FEM of the sample and contacts
4) Define a stimulation/measurement protocol coherent with the experiment
5) Chose a forward solver to calculate the $R_p$ transresistances at each iteration
6) Chose an inverse solver and **M** to define and minimize the functional (4.3)
7) Set the regularization parameter λ
8) Evaluate $R_S = 1/\langle\sigma(x,y)\rangle$, where $\langle\ .\ \rangle$ represent the spatial average.

## 4.3.  Bibliography

[1]     A. Cultrera, *et al*. "Mapping the conductivity of graphene with Electrical Resistance Tomography," *Sci. Rep.*, vol. 9, no. 1, pp 10655, 2019.

[2]     M. Cheney, D. Isaacson, and J. C. Newell, "Electrical Impedance Tomography," SIAM Review, vol. 41, no. 1, pp. 85–101, 1999.

[3]     L. Borcea, "Electrical Impedance Tomography," Inverse problems, vol. 18, no. 6, R99, 2002.

[4]     A. Adler and W. R. Lionheart, "EIDORS: towards a community-based extensible software base for EIT," in 6th Conf. on Biomedical Applications of Electrical Impedance Tomography, 1–4, 2005.

[5]     A. Cultrera and L. Callegaro, "Electrical Resistance Tomography of conductive thin films," IEEE Trans. Instrum. Meas., vol. 65, no. 9, pp. 2101–2107, 2016.

[6]     C. Melios, C. E. Giusca, V. Panchal, and O. Kazakova, "Water on graphene: Review of recent progress," 2D Materials, vol. 5, no. 2, 2018.

[7]     D. C. Dobson and F. Santosa, "Resolution and stability analysis of an inverse problem in electrical impedance tomography: dependence on the input current patterns," SIAM J. Appl. Math., vol. 54, no. 6, pp. 1542–1560, 1994.

[8]     X. Chen et al., "Breakdown current density of CVD-grown multilayer graphene interconnects*," IEEE Electr. Device L.*, vol. 32, no. 4, pp. 557-559, 2012.

[9]     X. Chen, D. H. Seo, S. Seo, H. Chung, and H. S. P. Wong, "Graphene interconnect lifetime: a reliability analysis," IEEE Electr. Device L., vol. 33, no. 11, pp. 1604 - 1606 , 2012.

[10]    M. Vauhkonen, D. Vadasz, P. A. Karjalainen, E. Somersalo, and J. P. Kaipio, "Tikhonov regularization and prior information in electrical impedance tomography," IEEE T.  Med. Imaging, vol. 17, no. 2, pp. 285–293, 1998.

[11]    L. Elden, "Algorithms for the regularization of ill-conditioned least squares problems." BIT Numer. Math., vol. 17, no 2, pp 134–145, 1977.

[12]    L. Kaufman and A. Neumaier, "PET regularization by envelope guided conjugate gradients," IEEE Trans. Med. Imaging, vol. 15, no.3, pp 385–389, 1996.

[13]    P. C. Hansen, "Regularization tools: a MATLAB package for analysis and solution of discrete ill-posed problems," Numer. Algorithms, vol. 6, no. 1, pp. 1–35, 1994.

# 5 Measurements with the coplanar waveguide method

## 5.1 General

This part of the protocol establishes a standardized method to determine the key control characteristic sheet resistance $R_S$ using the Co-Planar Waveguide (CPW) method. The method is applicable for graphene grown by CVD on or transferred to quartz substrates or other insulating materials as well as graphene on silicon carbide. The measurement range for the sheet resistance of the graphene layer shall be in the nominal range of $10^2 - 10^4 \ \Omega/\text{sq}$.

The measurement can be performed in two configurations:

1) Series configuration: graphene constitutes the central line of the CPW.

2) Shunt configuration: graphene is the CPW dielectric.

Which method is preferable is determined by the graphene conductivity $\sigma$.

### 5.1.1 Measurement principle

A coplanar waveguide consists of a strip of thin metallic film on the surface of a dielectric slab with two ground electrodes running adjacent and parallel to the strip. Practical applications of the coplanar waveguides have been experimentally demonstrated by measurements on resonant isolators and differential phase shifters fabricated on low-loss dielectric substrates with high dielectric constants [1]. Low-loss dielectric substrates with high dielectric constants may be employed to reduce the longitudinal dimension of the integrated circuits because the characteristic impedance of the coplanar waveguide is relatively independent of the substrate thickness. The measurement of the S-parameters by means of vector network analysers (VNA) allows to recover the ac conductivity of graphene when it is made part of the CPW itself [2,3].

### 5.1.2 Sample preparation method

Graphene CPW samples are prepared by geometrization (typically by lithography) and deposition of metal contacts. The contacts are shaped as a CPW. Two different configurations of samples are possible (see Fig. 5.1 and Sec. 5.2.2), they can be realised using the same lithographic methods. The graphene to be measured constitutes either the whole or a part of the central conductor of the CPW (in the "series" geometry) or fills the spacing between the CPW lines (in the "shunt" geometry). An example of lithographic process is reported in Ref. [2].

### 5.1.3 Description of measurement equipment / apparatus

CPW is a two-port rf network. The instrument that allows to measure the electrical parameters of a n-port network in a wide frequency range is the Vector Network Analyzer (VNA). The basic architecture of a VNA involves a signal generator, one or more detectors, directional couplers and switches. A test signal, which is typically a sinewave swept in frequency in a given bandwidth, is applied to one port (either port 1 or port 2); reflected signal by the same port and the transmitted signal to the other port are measured in magnitude and phase. In the S-parameter representation, the properties of the network being measured are given as a scattering parameter matrix $S = [S_{11}, S_{12}; S_{21}, S_{22}]$. $S$ is a complete representation of the (linear) two-port network. Other representations of a network are the impedance matrix $Z$ and the admittance matrix $Y$. All representations are linked through matrix mathematical expressions.

Standard VNA is provided with coaxial ports. The measurement on planar devices as a CPW is performed with a probe station, a mechanical device which allows to position and align two mechanical probes on the CPW. Each probe converts the coaxial transmission line connected to the VNA to a planar transmission line compatible with CPW geometry. The most common probe has a Ground-Signal-Ground geometry which adapts to the CPW geometry.

## 5.1.4 Calibration standards

VNA calibration is the process of determining the VNA systematic measurement errors, which are called error coefficients. The process is based on a VNA error model. A VNA error model relates the values indicated by the VNA (raw S-parameters) to the S-parameters of the device connected to the test port of the VNA. Part of this mathematical relationship are the unknown error coefficients. During VNA calibration, a set of known calibration standards is measured, and the error coefficients are determined. They can then be used to transform the raw S-parameters of a DUT measurement into corrected S-parameters, a step which is usually being referred to as VNA error correction.

Wafer calibration standards are provided to calibrate the VNA together with the probes and define a reference plane for the measurements at the probe tips.

Several types of wafer calibration standards are available: SOLT (short - open - load - thru); an alternative to SOLT is the SOLR (short - open - load - reciprocal, also known as Unknown Thru) calibration. TRL (through - reflect - line) is employed in the higher frequency range.

## 5.1.5 Ambient conditions during measurement

Being a carbon monolayer, graphene electrical properties are strongly affected by ambient conditions, in particular the humidity [2]. The ambient conditions during the whole duration of the measurement must be monitored and recorded. Typical ambient conditions are those of electrical calibration laboratories, T = (23 ± 1) °C, RH = (50 ± 4) % but can be chosen from the table 1 in of IEC 60068-1.

## 5.2    Measurement procedure

The electrical properties of graphene modify the characteristic impedance and the propagation constant of the CPW. The CPW S-matrix is measured with the VNA. A mathematical model is employed to recover, from the S matrix and the geometrical dimensions of the CPW, the conductivity and permittivity of graphene.

## 5.2.1 Calibration of measurement equipment

VNA and probe station calibration are performed with a wafer calibration kit, which includes several microwave structures having a certified S-matrix. Modern VNA software applications are available that guide the operator through the VNA calibration steps and the subsequent measurement.

Manufacturer operation manual should be followed for compatibility between the specific calibration kit and the VNA probe of choice. For example, when using calibration kits with aluminium metallization, the probe pads must break through the aluminium oxide and therefore hard-metal (nickel or tungsten) probe tips are required.

The calibration is part the measurement procedure. The geometry of the calibration kit structures employed in the calibration is dependent on the geometry of the CPW under measurement.

## 5.2.2  Detailed protocol of the measurement procedure

### Single CPW

When the sample includes a single CPW, the VNA must be calibrated in order to define the reference planes at the level of the CPW ports.

- The probes are mounted on the probe station holders.
- The sample (or, in the calibration phase, the calibration kit) is positioned on the probe station under the optical microscope, focusing on the sample (the probes tips if in the field of view, should appear out of focus)
- The probes are positioned above the sample at a significant vertical distance, to avoid lateral scratching -- which would damage both the probes and the sample. Once positioned above the pads, the probes are slowly lowered on the sample. While approaching the sample, the probes enter the focus plane of the microscope field of view. Lateral adjustment of the position can be made in order to land the probe on the desired position on the sample pads. The contact between the probes and the sample can be checked by looking at the VNA response.
- The calibration procedure is performed. The sequence depends on the geometry and the calibration method implemented in the VNA software. The calibration can involve one or both ports, depending on the measurement chosen (one- or two-port scattering parameters).
- The probes are raised, and the calibration kit is replaced with the sample. The probe approach sequence is repeated.
- The measurement starts. The VNA performs a sequence of excitation and measurements at the two ports and recovers the individual components of the scattering parameter matrix.
- At the end of the measurement, the measurement software displays the frequency dependence of the scattering parameter matrix elements with the desired representation (bode diagram, Smith chart).

### Multiple CPW with different lengths

The sample may be provided with several CPW lines having different lengths (with all other geometrical parameters kept equal), see Fig. 5.2. If this set of structures is available, the material properties can be extracted without the need to perform a VNA calibration, since an accurate determination of the VNA reference planes is not necessary and raw measurement data can be used as input of the measurement model.

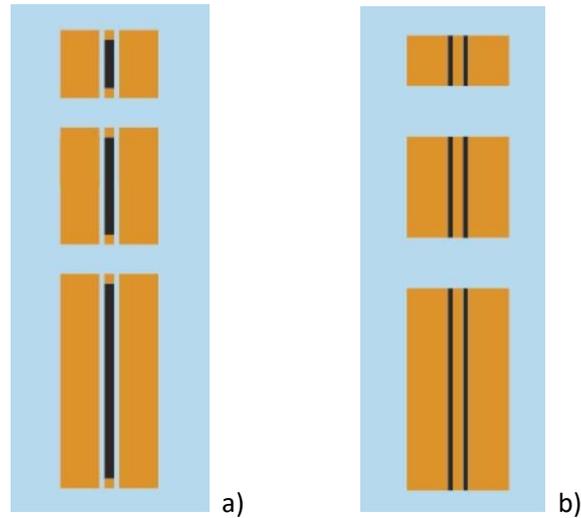

**Figure 5.1 – Possible CPW configurations of different lengths for the characterisation of graphene.** *Series* (a) and *shunt* (b) configurations are shown, graphene in black, metal in yellow, quarts substrate in light blue.

### 5.2.3  Measurement accuracy

The EURAMET Guide on calibration of VNA [4] gives lists the quantities which mainly affect the measurement of S-parameters accuracy:

- Characterization of calibration standards
- VNA noise floor and trace noise
- VNA non-linearity
- VNA drift
- Isolation (cross-talk)
- Test port cable stability
- Connection repeatability

When performing measurements on individual CPWs the measurement accuracy is affected by the determination of the reference planes in the course of the calibration phase: the distance between the reference planes should match the distance between the physical landing points of the probe fingers, and the CPW should not extend beyond these planes. Hence, an accurate positioning of the probes at the CPW ports is essential.

For the specific case of multiple-length CPW (see 5.2.2), since no calibration is performed, the absolute distance between reference planes is not determined, and only differences in the reference plane spacing for the measurements on the various CPW lengths is considered in the measurement model.

### 5.2.4  Data analysis / interpretation of results

Several CPW calibration and measurement methods are available. As an example, we give here a summary of the measurement model that can be employed for the multiple-length CPW described in 5.2.2, see [5].

Consider the case of two CPW lines of lengths $L_1$ and $L_2$ (and otherwise identical). The two transmission matrices T($L_1$) and T($L_2$) can be computed from the measured scattering parameter matrices S($L_1$) and S($L_2$) [5].

The transmission matrix of a CPW with equivalent length $L = L_2 - L_1$ can be computed with the cascade rule:
$P_1$ --- $L_1$ ---- $L$ ---- $P_2$,

$$\mathbf{T}(L_2) = \mathbf{T}(L) \bullet \mathbf{T}(L_1), \qquad (5.1)$$

hence

$$\mathbf{T}(L) = \mathbf{T}(L_2) \bullet \mathbf{T}^{-1}(L_1). \qquad (5.2)$$

The complex propagation constant of a CPW of length $L$ is $\gamma = \alpha + j\beta$, where $\alpha$ is the line attenuation (dependent on the resistivity) and $\beta$ is the line phase constant (a measure of the speed of the electromagnetic waves, related to the refractive index) [1, 6].

The quantity $\gamma$ can be computed from the eigenvalues of $\mathbf{T}(L)$, see sec. IV of [1] for the extended solution of the eigenvalue problem.

Once $\gamma$ is computed from $\mathbf{T}(L)$, $\alpha$ can be obtained. The relation between $\alpha$ and the conductivity of the graphene included in the DUT can be retrieved from the relations reported in [7].

## 5.3   Bibliography

[1]   G.F. Engen and C.A. Hoer, "Thru-Reflect-Line: an improved technique for calibrating the dual six-port automatic network analyzer", IEEE *Trans. Microwave Theory Tech.,* vol. MTT-27, pp. 987-993, 1979.

[2]   H. Skulason, *et al.*, "110 GHz measurement of large-area graphene integrated in low-loss microwave structures," Appl. Phys. Lett., vol. 99, no. 15, pp. 153504, 2011.

[3]   M. Dragoman, *et al.*, "Coplanar waveguide on graphene in the range 40 Mhz–110 Ghz," Appl. Phys. Lett., vol. 99, no. 3, pp. 033112, 2011.

[4 ]   Available online at:
https://www.euramet.org/Media/news/I-CAL-GUI-012_Calibration_Guide_No._12.web.pdf

[5]   G. Ghione, *et al.*, "Microwave modeling and characterization of thick coplanar waveguides on oxide-coated lithium niobate substrates for electrooptical applications," IEEE *Trans. Microwave Theory Tech.* vol. 47, no. 12, pp. 2287–2293, 1999.

[6]   R.B. Marks, "A multiline method of network analyser calibration", IEEE *Trans. Microwave Theory Tech.,* vol. MTT-39, pp. 1205-1215, 1991

[7]     B. Benarabi, *et al.*, "Microwave characterization of electrical conductivity of composite conductors by half-wavelength coplanar resonator," Progress in Electromagnetics Research, vol. 60, pp. 73–80, 2016.

## Conclusions

Using this guide, users with previous characterisation expertise within industry can more reliably, quantitatively and comparably measure the electrical properties of commercially supplied graphene.

This guide forms the basis for future international standards in this area, specifically, standards currently under development within IEC/TC 113 'Nanotechnologies' for the characterisation of the electrical properties of graphene. This will lead to the continual improvement in the measurement of the electrical properties of graphene, as well as reveal any reproducibility issues in performing these graphene measurements in different laboratories across the world.

## Acknowledgements

The work is part of the European project "GRACE — Developing electrical characterisation methods for future graphene electronics", code 16NRM01. This project has received funding from the European Metrology Programme for Innovation and Research (EMPIR) programme co-financed by the Participating States and from the European Union's Horizon 2020 research and innovation programme.

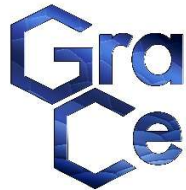